\documentclass[prc,preprint,superscriptaddress,showkeys,showpacs,
   preprintnumbers,amsmath,amssymb,aps,floatfix]{revtex4-1}
\usepackage{graphicx}
\usepackage{epsfig}
\usepackage{amsfonts}
\usepackage{amsmath}
\usepackage{amssymb}
\usepackage[sort&compress]{natbib}
\usepackage{dcolumn}
\usepackage{multirow}
\usepackage{bigstrut}
\usepackage{type1cm}
\usepackage{placeins}
\usepackage{rotating}
\usepackage{slashed}

\newcommand{\Ip}{\hat{\mathrm{I}}}

\newcommand{\se}{\hat{s}_e}

\newcommand{\ccr}{\chi_r}
\newcommand{\cci}{\chi_i}

\begin{document}

\preprint{\today}

\title{Lorentz violation in neutron and allowed nuclear $\mathbf\beta$ decay}

\author{J. P. Noordmans}
\author{H. W. Wilschut}
\author{R. G. E. Timmermans}

\affiliation{KVI, University of Groningen, Zernikelaan 25,
                   NL-9747 AA Groningen, The Netherlands}

\date{\today}
\vspace{3em}

\begin{abstract}
\noindent
\begin{description}
\item[Background]
The search for violations of Lorentz invariance is nowadays motivated by attempts to unify the
Standard Model of particle physics with general relativity. Such theories of ``quantum gravity''
predict Lorentz-violating signals that could be detected in low-energy precision experiments.
In this context, Lorentz invariance has been tested poorly in the weak interaction.
\item[Purpose]
We explore the possibility that the weak interaction violates Lorentz, and in particular rotational, invariance in
neutron and allowed nuclear $\beta$ decay.
\item[Method]
A broad class of Lorentz-violating effects is considered in an effective field theory approach, wherein the
standard propagator of the $W$-boson acquires an additional Lorentz-violating tensor.
\item[Results]
The general decay rate for allowed $\beta$ decay that incorporates the modified $W$-boson propagator is derived.
The resulting Lorentz-violating signals are discussed for the different types of $\beta$-decay transitions: Fermi, Gamow-Teller,
and mixed. We study the implications of our formalism for dedicated $\beta$-decay experiments. We give a short overview
of the few relevant experiments that have been performed in the past or are ongoing.
\item[Conclusions]
Our work provides a general theoretical framework that should be used for designing and interpreting $\beta$-decay
experiments that search for Lorentz violation. In particular, it determines the kind of experiments that are
necessary to probe different parameters that quantify Lorentz violation, and it establishes their sensitivity.
\end{description}
\end{abstract}
\maketitle

\section{Introduction}
Symmetries of spacetime, and Lorentz and CPT symmetry in particular, are at the root of our present understanding of nature~\cite{Ein05,Wig39,Wei95}. The best theories of nature we have today, the Standard Model (SM) of particle physics and general relativity, conserve these symmetries. Unification of these theories is a major goal of theoretical high-energy physics. Different approaches have been put forward, but none of these is completely satisfactory. One of the problems is that the energies relevant for the unification of quantum mechanics and general relativity are far out of reach of present-day experiments.

An opportunity may lie in the possibility that Lorentz and/or CPT symmetry are broken at high energy. In some candidate theories of quantum gravity, mechanisms of Lorentz violation have been identified (see {\it e.g.} Refs. \cite{Kostelecky:1988zi, Ellis:1999yd, Burgess:2002tb, Gambini:1998it}). The phenomenological consequences at low energy of such a breakdown of Lorentz symmetry have been studied over the last few decades. It has become clear that it is possible to constrain violations of Lorentz symmetry with enormous precision \cite{Kostelecky2011}---to such precision, in fact, that the relevant effects have been termed ``windows on quantum gravity.'' Since the underlying theory of quantum gravity is unknown and no compelling evidence of Lorentz violation has been found, the approach one should arguably take to investigate the possibility of Lorentz violation is that of an effective field theory that encompasses all possible Lorentz-violating operators. In this respect, the Lorentz-violating Standard Model Extension (SME) \cite{Colladay1997,Colladay1998a,Colladay1998b} has become an influential and indispensable tool in the search for Lorentz violation.

The first discovery of a violation of a presumed (discrete) spacetime symmetry was that of parity violation in nuclear $\beta$ decay \cite{Lee56,Wu:1957my}. This discovery led to the understanding that the weak interaction has a ``$V$$-$$A$'' structure, {\it i.e.} it is mediated by the $W$-boson that couples only to left-handed fermions. Experiments on neutron and allowed nuclear $\beta$ decay have contributed much to the development of the SM in the 1960s and 1970s. In the last decade,  we witnessed a renaissance of nuclear $\beta$ decay due to the novel experimental techniques of laser cooling and atom trapping, which enabled a new generation of $\beta$-decay experiments, wherein the momentum of the recoiling daughter nucleus could be detected and new observables became accessible. The motivation for these modern experiments is to search for physics {\it beyond} the SM, in particular so-called non-$V$$-$$A$ (right-handed vector, scalar, and tensor) currents~\cite{Her01,Sev06,Sev11}. In this work, we want to study in a general way the possibility that the weak interaction violates Lorentz symmetry, and in particular rotational invariance~\cite{Lee56}, in $\beta$ decay. Such Lorentz violation would give rise to unique signals with no SM ``background,'' which, even when tiny, could be experimentally detectable. 

We will consider a general Lorentz-violating correction to the propagator of the $W$-boson and calculate its effect on the ``allowed'' $\beta$-decay rate of a nucleus. The Lorentz-violating propagator at low energies that we will use is given by
\begin{equation}
\left\langle W^{\mu+}(p)W^{\nu-}(-p)\right\rangle = \frac{-i(g^{\mu\nu}+\chi^{\mu\nu})}{M_W^2} \ ,
\label{wpropagator}
\end{equation}
where $g^{\mu\nu}$ is the Minkowski metric and $\chi^{\mu\nu}$ is a general Lorentz-violating (complex, possibly momentum-dependent) tensor. Calculations comparable to ours have been done in Refs. \cite{Nielsen:1982kx, Huerta:1983hg}, however, these were restricted to a real, traceless, and rotationally invariant tensor. The effect on $\beta$ decay of Lorentz violation in neutrinos is considered in Ref. \cite{Diaz2013}.

At this point, we place no restrictions on $\chi^{\mu\nu}$ other than tracelessness. A trace part of the tensor does not violate Lorentz symmetry and can be absorbed in the coupling constant. The result we get in terms of $\chi^{\mu\nu}$ is quite general and can be used to read off results for any SME parameter that shows up in the propagator of the $W$-boson. Effects that we are missing at tree level by restricting ourselves to Eq. ~\eqref{wpropagator} come from the Lorentz-violating contributions to the free Dirac equation, which modify the asymptotic in- and out-states. For the nucleon, electron, and positron, they can be constrained better by QED observables, while we focus on the weak interaction, for which much less is known in the context of Lorentz violation. Exceptions to the expectation of higher sensitivity of QED tests to free-fermion parameters are parameters that can be removed from the free theory by field redefinitions \cite{Colladay:2002eh} and show up in the interaction with $W$- or $Z$-bosons only. A complete list of parameters, or combinations thereof, that show up as $\chi^{\mu\nu}$ in Eq.~\eqref{wpropagator} lies beyond the scope of this paper.

The organization of this paper is as follows. In Sec. \ref{sec:general} we derive in a general way the rate of allowed $\beta$ decay including the effects of Lorentz violation. In Sec. \ref{sec:chicontrib} we analyze the structure of the tensor $\chi^{\mu\nu}$ that supplements the standard propagator of the $W$-boson, and argue that it represents a broad class of Lorentz-violating effects. In Sec. \ref{sec:discussion} we study the implications of our formalism for experiments. We discuss the possible signals in the different types of allowed $\beta$ decays, and we give a short overview of the few relevant experiments that have been performed or are ongoing. Three Appendices are devoted to intermediate or detailed results of the derivations or discussions in the main text.

\section{Allowed $\beta$-decay rate including Lorentz violation} \label{sec:general}
Our calculation parallels the standard calculations of (polarized) $\beta$ decay in the conventional $V$$-$$A$ framework. We will discuss the derivation of the $\beta$-decay rate to make clear where the Lorentz violation enters and how it modifies the standard results. We follow the conventional notation for $\beta$-decay calculations \cite{Jackson:1957zz, Jackson1957206, Ebel1957213,konopinski1966theory}. We use units such that $\hbar=c=1$, our metric is ``mostly minus,'' and $\gamma^5 \equiv i\gamma^0\gamma^1\gamma^2\gamma^3$. The effective $\beta$-decay Hamiltonian density that follows from $V$$-$$A$ theory, including the Lorentz-violating $W$-boson propagator Eq.~\eqref{wpropagator}, is given by
\begin{equation}
\mathcal{H}_\beta = (g_{\rho\sigma}+\chi_{\rho\sigma})\left[\bar{\psi}_p(x)\gamma^\rho(C_V+C_A\gamma^5)\psi_n(x)\right]\left[\bar{\psi}_e(x)\gamma^\sigma(1-\gamma^5)\psi_\nu\right(x)]+h.c. \ ,
\label{betahamiltonian}
\end{equation}
where $C_V$ and $C_A$ are real constants that determine the relative amplitude of the vector and axial-vector interaction, and the gamma matrices are in the Dirac basis; $h.c.$ denotes hermitian conjugation. 

We make two approximations that are customary in $\beta$-decay calculations. In the first place, we evaluate the lepton current at the nuclear center, because the de Broglie wavelengths of the leptons are much larger than the nuclear radius and the current is practically constant over that range. This implies that the leptons take away zero orbital angular momentum. Secondly, since the nuclei are nonrelativistic, the ``small'' lower two components of the nuclear wavefunction can be neglected. These two approximations comprise {\it allowed} $\beta$ decay. With these approximations, we can write the squared matrix element that follows from Eq.~\eqref{betahamiltonian} as
\begin{equation}
|\mathcal{M}|^2 = \left|C_V\left\langle 1 \right\rangle J^0_{\mp} - C_A \left\langle \boldsymbol{\sigma}\right\rangle \cdot \bf{J}_{\mp} \right|^2 \ ,
\label{squaredmatrixelement1}
\end{equation}
where $J_{\mp}$ is the lepton current for $\beta^{\mp}$ decay evaluated at the nuclear center, into which we absorbed the Lorentz violation,
\begin{subequations}
\begin{eqnarray}
J_{-}^\rho & = & (g^{\rho\sigma}+\chi^{\rho\sigma})\bar{\psi}_{e^-}(0)\gamma_\sigma(1-\gamma^5)\psi_{\bar{\nu}}(0) \ ,
\label{leptoncurrenta} \\
J_{+}^\rho & = & (g^{\rho\sigma}+\chi^{\rho\sigma*})\bar{\psi}_\nu(0)\gamma_\sigma(1-\gamma^5)\psi_{e^+}(0) \ ,
\label{leptoncurrentb}
\end{eqnarray}
\end{subequations}
and $\left\langle 1 \right\rangle = \left\langle f|1|i\right\rangle$ and $\left\langle \boldsymbol{\sigma} \right\rangle = \left\langle f|\boldsymbol{\sigma}|i\right\rangle$ abbreviate the transition matrix elements of the nucleus.

The charged lepton wavefunction is not given by the plane-wave solution of the free Dirac equation, since the $\beta$ particle feels the positive charge of the daughter nucleus after the decay. We take the spinor for allowed $\beta$ decay from Ref.~\cite{springerlink:10.1007/BF01340460} (our normalization differs by a factor $\sqrt{2E_e}$); it reads
\begin{equation}
\psi^s_{e^-}(r\rightarrow 0) = N(Z)\sqrt{E_e+m_e}\left(\begin{array}{c}\eta^s \\ M \frac{\boldsymbol{\sigma}\cdot\bf{p}}{E_e+m_e}\eta^s\end{array}\right) \ ,
\label{electronwavefunction}
\end{equation}
where
\begin{equation}
|N(Z)|^2 = \frac{E_e+\gamma m_e}{E_e+m_e}F(E_e,Z), \qquad \gamma = \sqrt{1-(Z\alpha)^2} \ ,
\end{equation}
and
\begin{equation}
M = \frac{E_e+m_e}{E_e+ \gamma m_e}\left(1+i\frac{Z\alpha m_e}{|\bf{p}|}\right) \ .
\end{equation}
Here $E_e$, $m_e$, and $\bf{p}$ are the electron energy, mass, and momentum, respectively, and $\eta^s$ is a Pauli spinor. $Z$ is the atomic number of the daughter nucleus, $\alpha$ the fine-structure constant, and $F(E_e,Z)$ the usual Fermi function, which is in essence the probability to find an electron in the interior of the nucleus relative to the probability to find it at the same position without a nucleus present. The spinor of the positron is the charge conjugate of Eq.~\eqref{electronwavefunction}, which amounts to
\begin{equation}
\psi^s_{e^+}(r\rightarrow 0,Z) = i \gamma^2 (\psi^s_{e^-}(r\rightarrow 0,-Z))^* \ .
\label{positronwavefunction}
\end{equation}
The wavefunction of the neutrino is just the solution of the free massless Dirac equation.

Since there is no Lorentz violation in the hadronic current, the evaluation of nuclear transition matrix elements proceeds as usual through the Wigner-Eckart theorem~\cite{rose1995elementary}, which allows one to write the matrix elements of the operators $1$ and $\boldsymbol{\sigma}$, given in terms of their spherical tensor components, as a product of Clebsch-Gordan coefficients and matrix elements that are independent of the spin projection quantum number $m$. The result is that we can write the squared matrix element Eq.~\eqref{squaredmatrixelement1} as
\begin{eqnarray}
|\mathcal{M}|^2& = &\left\{C_V^2 \left\langle 1 \right\rangle^2 \delta_{jj'}|J^0|^2 + \tfrac{1}{3}C_A^2\left\langle\sigma\right\rangle^2 (|J_{+1}|^2 + |J_{-1}|^2 + |J_z|^2) \right. \notag \\
 && + \tfrac{1}{2}C_A^2\left\langle\sigma\right\rangle^2 \Lambda^{(1)}(|J_{+1}|^2 - |J_{-1}|^2) + \tfrac{1}{2}C_A^2\left\langle\sigma\right\rangle^2 \Lambda^{(2)}(|J_{+1}|^2 + |J_{-1}|^2 - 2 |J_z|^2) \notag \\
 &&  \left. - C_V C_A \left\langle 1 \right\rangle \left\langle \sigma \right\rangle \delta_{jj'}\Lambda_z(J^0 J^*_z+J_z J^{0*})\right\} \ ,
\label{squaredmatrixelement2}
\end{eqnarray}
where the space components of the lepton current are now given in spherical coordinates as
\begin{equation}
J_{\pm 1} = \mp\tfrac{1}{\sqrt{2}}(J^1 \pm i J^2) \ , \qquad \mathrm{and} \qquad J_z = J^3 \ ;
\end{equation}
$j$ and $j'$ denote the initial and final nuclear spin, respectively, $\left\langle 1\right\rangle$ and $\left\langle \sigma \right\rangle$ are the reduced matrix elements independent of the spin projection, while the $\Lambda$ coefficients come from combinations of Clebsch-Gordan coefficients. They are given by
\begin{equation}
\Lambda^{(1)} =\left\{\begin{array}{ll}
\frac{\left\langle m \right\rangle}{j} & (j' = j-1) \\
\frac{\left\langle m \right\rangle}{j(j +1)} & (j' = j) \\
\frac{-\left\langle m \right\rangle}{j + 1} & (j'=j+1)
\end{array}
\right., \qquad
\Lambda^{(2)} = \left\{\begin{array}{ll}
\frac{\left\langle m^2 \right\rangle - \frac{1}{3}j(j + 1)}{j(2j - 1)} & (j' = j-1) \\
\frac{-\left\langle m^2 \right\rangle + \frac{1}{3}j(j + 1)}{j(j + 1)} & (j' = j) \\
\frac{\left\langle m^2 \right\rangle - \frac{1}{3}j(j + 1)}{(j+1)(2j + 3)} & (j'=j+1)
\end{array}
\right. \ ,
\end{equation}
and 
\begin{equation}
\Lambda_z = \frac{\left\langle m \right\rangle}{j}\sqrt{\frac{j}{j+1}} \ .
\end{equation}
Here the notation $\left\langle m \right\rangle$ and $\left\langle m^2 \right\rangle$ means that $m$ and $m^2$ must be averaged (incoherently) over the populations of the states $m=\pm j,\pm(j-1),\ldots$ For a completely polarized source this implies that $\left\langle m \right\rangle = j$. We denoted the $m=0$ component of the current in spherical coordinates by $J_z$ to distinguish it from the time component of the current $J^\mu$.

The first term in Eq.~\eqref{squaredmatrixelement2} corresponds to a Fermi transition, which has $\Delta j = j-j' = 0$. This Fermi term is isotropic, and in particular independent of the nuclear polarization. The terms including the factor $C_A^2$ give Gamow-Teller transitions with $\Delta j = 0,\pm 1$. The first of these Gamow-Teller terms has $\Delta j = 0$ and is also isotropic. The second term proportional to $C_A^2$ is ``first-order anisotropic,'' characterized by the fact that it is proportional to $\Lambda^{(1)}\propto\left\langle m\right\rangle / j$. This term gives transitions with $\Delta j = 0,\pm 1$. The third Gamow-Teller term, following $\Lambda^{(2)}$, is ``second-order anisotropic'' and also has $\Delta j = 0,\pm 1$. Finally, the last term proportional to $C_V C_A$ causes a so-called ``mixed'' transition. It is first-order anisotropic and gives $\Delta j = 0$. All anisotropic terms average to zero for randomly oriented nuclei.

Up to this point, the calculation is identical to the Lorentz-symmetric case, except for the presence of the Lorentz violation in the lepton current Eqs.~\eqref{leptoncurrenta} and \eqref{leptoncurrentb}. To work out Eq.~\eqref{squaredmatrixelement2} we need products of different components of the lepton current, by evaluating the general product $J^\mu J^{\nu*}$. Since the electron and positron spinors Eqs.~\eqref{electronwavefunction} and \eqref{positronwavefunction} can be written as
\begin{subequations}
\begin{eqnarray}
   \psi^s_{e^-}(0) &=& \frac{N(Z)}{2}\left[(1+\gamma^0)+M(1-\gamma^0)\right]u^s(p) \ , \\  
   \psi^s_{e^+}(0) &=& \frac{N^*(-Z)}{2}\left[(1-\gamma^0)+M(1+\gamma^0)\right]v^s(p) \ ,
\end{eqnarray}
\end{subequations}
with $u^s(p)$ and $v^s(p)$ the free Dirac spinors for the electron and positron, respectively, it is straightforward to show that for the electron
\begin{subequations}
\begin{equation}
\sum_{\nu\;\mathrm{spin}} J_-^\mu J_-^{\nu*} = F(E_e,Z) (g^{\mu\rho}+\chi^{\mu\rho})(g^{\nu\sigma}+\chi^{\nu\sigma*})\mathrm{Tr}\left[ \slashed{P}_-\gamma_\rho\slashed{k}\gamma_\sigma(1-\gamma^5)\right] \ ,
\label{currenttraceelec}
\end{equation}
while for the positron
\begin{equation}
\sum_{\nu\;\mathrm{spin}} J_+^\mu J_+^{\nu*} = F(E_e,-Z) (g^{\nu\rho}+\chi^{\nu\rho})(g^{\mu\sigma}+\chi^{\mu\sigma*})\mathrm{Tr}\left[ \slashed{P}_+\gamma_\rho\slashed{k}\gamma_\sigma(1-\gamma^5)\right] \ ,
\label{currenttracepos}
\end{equation}
\end{subequations}
where the neutrino four-momentum is $k^\mu$. We sum over the neutrino polarizations in Eqs.~\eqref{currenttraceelec} and \eqref{currenttracepos}, because these are unlikely to be measured in the foreseeable future. $P^\mu_\mp$ is given by
\begin{subequations}
\begin{eqnarray}
   P_{\mp}^0 &=& E_e \mp {\bf{p}\cdot\hat{s}_e} \ , \\ 
   P_{\mp}^i &=& \left(1\mp\frac{{\bf(p\cdot\hat{s}_e)}(E_e-\gamma m_e)}{|{\bf p}|^2}\right)p^i \mp m_e\gamma\hat{s}_e^i \mp \sqrt{1-\gamma^2} m_e({\bf \hat{\bf p}}\times{\bf \hat{s}_e})^i \ ,
\label{defP}
\end{eqnarray}
\end{subequations}
with $i$ running over the spatial indices $1,2,3$, and ${\bf \hat{s}_e}$ the unit vector in the direction of the polarization of the $\beta$ particle. The upper (lower) signs correspond to the electron (positron). Notice that $P_{\mp}^\mu$ does not transform as a four-vector, because the Coulomb corrections are included in a boost non-invariant way in Ref.~\cite{springerlink:10.1007/BF01340460}. At present, this does not concern us, since we will mainly consider the effects of rotations. Neglecting the Coulomb corrections $(Z\alpha \rightarrow 0)$ gives $P_{\mp}^\mu \rightarrow p^\mu \mp m_e s_e^\mu$, with $s^\mu$ the conventional spin four-vector.

It now becomes a simple matter of trace technology to calculate the different products of lepton currents needed in Eq.~\eqref{squaredmatrixelement2}. Although this calculation is straightforward, the results are rather lengthy and not very illuminating. Their exact form is given in Appendix \ref{lepcurrentsapp}. Substituting the results for the currents in Eq.~\eqref{squaredmatrixelement2} and remembering that the differential decay rate is given by
\begin{equation}
dW = \frac{\delta(E_e + E_\nu - E_0)}{(2\pi)^5 2E_e 2 E_\nu}\sum_{\nu\;\mathrm{spin}}|\mathcal{M}|^2 d^3p\,d^3k \ ,
\end{equation}
where $E_0$ is the total energy available in the decay, we find the general Lorentz-violating result for allowed $\beta$ decay, including Coulomb corrections and to first order in $\chi^{\mu\nu}$, given by
\begin{multline}
dW= \frac{1}{(2\pi)^5}d^3p\,d^3k\,\delta(E_e + E_\nu - E_0) F(E_e,\pm Z)\xi  \\
\times \; \left\{\left(1\mp\frac{\bf{p}\cdot\se}{E_e}\right)\left[\tfrac{1}{2}\left(1+B\frac{{\bf k \cdot \hat{I}}}{E_\nu}\right) + t +\frac{\boldsymbol{w}_1\cdot{\bf k} }{E_\nu} + \boldsymbol{w}_2\cdot {\bf \hat{I}} + T_1^{km}\Ip^k \Ip^m + \frac{T_2^{kj}\Ip^k k^j}{E_\nu} + \frac{S_1^{kmj}\Ip^k \Ip^m k^j}{E_\nu} \right]\right.\\
+\left(\left(1\mp\frac{(E_e-\gamma m_e)({\bf p\cdot \hat{s}_e})}{E_e^2-m_e^2}\right)\frac{p^l}{E_e}\mp\frac{\gamma m_e}{E_e}\hat{s}_e^l \mp \frac{m_e}{E_e}\sqrt{1-\gamma^2}({\bf \hat{p} \times \hat{s}_e})^l\right) \\
\left.\times\left[\tfrac{1}{2}\left(A-3c\frac{{\bf k\cdot\hat{I}}}{E_\nu}\right)\Ip^l + \tfrac{1}{2}(a+c)\frac{k^l}{E_\nu} + w_3^l + \frac{T_3^{lj}k^j}{E_\nu} + T_4^{lk}\Ip^k + S_2^{lmk}\Ip^m\Ip^k + \frac{S_3^{lmj}\Ip^m k^j}{E_\nu} + \frac{R^{lmkj}\Ip^m\Ip^k k^j}{E_\nu}\right]\right\} \ ,
\label{finalbetaresult}
\end{multline}
where ${\bf \hat{I}}$ is the nuclear polarization axis. The Latin indices run over the three spatial directions and summation over repeated indices is implied. The upper sign corresponds to $\beta^-$ decay, while the lower sign corresponds to $\beta^+$ decay. The different Lorentz-violating quantities are defined as
\begin{gather}
t = (a-\tfrac{1}{2}c)\chi_r^{00} \ , \notag \\
w_1^j = -x\chi_r^{0j} - \breve{g}(\tilde{\chi}_i^j-\chi_r^{j0}) \ , \qquad w_2^k = \breve{K}(\chi_r^{k0}-\chi_r^{0k})-\breve{L}\tilde{\chi}_i^k \ , \qquad w_3^l = x\chi_r^{0l}+\breve{g}(\chi_r^{l0} + \tilde{\chi}_i^l) \ , \notag \\
T_1^{km} = \tfrac{3}{2}c\chi_r^{km} \ , \qquad T_2^{kj} = \tfrac{1}{2}A\chi_r^{00}\delta^{jk} + \breve{L}(\chi_r^{jk}+\chi_i^{s0}\epsilon^{sjk})-\breve{K}(\chi_r^{kj}+\chi_i^{0s}\epsilon^{sjk}) \ , \notag \\
 T_3^{lj} = (x+\breve{g})\chi_r^{00}\delta^{lj}-(x\chi_i^{0s} + \breve{g}\chi_i^{s0})\epsilon^{sjl}-\breve{g}(\chi_r^{jl}+\chi_r^{lj}) \ , \notag \\ 
T_4^{lk} = \tfrac{1}{2}B\chi_r^{00}\delta^{lk}-\breve{L}(\chi_r^{lk} - \chi_i^{s0}\epsilon^{ksl})-\breve{K}(\chi_r^{kl}-\chi_i^{0s}\epsilon^{ksl}) \ , \notag \\
S_1^{kmj} = -\tfrac{3}{2}c(\chi_r^{k0}\delta^{mj}-\chi_i^{ms}\epsilon^{sjk}) \ , \qquad S_2^{lmk} = -\tfrac{3}{2}c(\chi_r^{m0}\delta^{kl}+\chi_i^{ms}\epsilon^{slk}) \ , \notag \\
S_3^{lmj} = \breve{L}\left(\chi_r^{l0}\delta^{jm}-\chi_i^{sl}\epsilon^{sjm}-\chi_r^{j0}\delta^{ml}+\tilde{\chi}_i^m \delta^{jl} - \chi_i^{sj}\epsilon^{lms}\right) \qquad\qquad  \notag \\ 
\qquad\qquad+ \breve{K}\left(\chi_i^{00}\epsilon^{ljm}-\chi_r^{0l}\delta^{jm}-\chi_r^{0j}\delta^{ml}+(\chi_r^{0m}+\chi_r^{m0})\delta^{jl}-\chi_i^{ms}\epsilon^{sjl}\right), \notag \\
R^{lmkj}= \tfrac{3}{2}c\left(\chi_i^{m0}\epsilon^{lkj} - \chi_r^{mk}\delta^{lj} + \chi_r^{ml}\delta^{kj} + \chi_r^{mj}\delta^{kl}\right) \ ,
\label{LVquantities}
\end{gather}
with the subscripts $r$ and $i$ denoting the real and imaginary parts of $\chi^{\mu\nu}$, respectively, $\tilde{\chi}^l = \epsilon^{lmk}\chi^{mk}$, and the different constants are given by
\begin{gather}
\xi = 2(C_V^2\left\langle 1 \right\rangle^2 + C_A^2\left\langle \sigma \right\rangle^2) \ , \qquad x = \frac{2C_V^2\left\langle 1 \right\rangle^2}{\xi} \ , \qquad y = \frac{2C_V C_A \left\langle 1 \right\rangle \left\langle \sigma \right\rangle}{\xi} \ , \qquad a = \tfrac{1}{3}(4x-1)\ , \notag \\
\qquad c = (1-x)\Lambda^{(2)} \ , \qquad A = \mp(1-x)\Lambda^{(1)}-2y\Lambda_z \delta_{jj'} \ , \qquad B = \pm(1-x)\Lambda^{(1)}-2y \Lambda_z \delta_{jj'} \ ,\notag \\
\breve{g} = \tfrac{1}{3}(1-x)(1+\tfrac{3}{2}\Lambda^{(2)}) \ , \qquad \breve{K}=-y\Lambda_z \delta_{jj'} \ , \qquad \breve{L} = \pm\tfrac{1}{2}(1-x)\Lambda^{(1)} \ ;
\label{defconstants}
\end{gather}
again, upper (lower) signs refer to $\beta^{-}(\beta^{+})$ decay. 

The nomenclature of the constants in Eq.~\eqref{defconstants} is such that the quantities without a ``breve'' occur in standard $\beta$ decay, while the ones with a ``breve''  are Lorentz violating. Capital letters signal terms which are first-order anisotropic and the other terms are in small script. Our expressions are compatible with the existing literature for the coefficients that occur in the  SM case as well (except that, for convenience, we absorbed a factor $\tfrac{1}{3}\Lambda^{(2)}$ in $c$). The $A$, for example, denotes the conventional ``Wu parameter'' that quantifies the correlation between the momentum of the $\beta$ particle and the polarization of the parent nucleus. Furthermore, we gave some thought to the order of the terms in Eq.~\eqref{finalbetaresult}. First of all, the second line contains terms that can experimentally be tested without referring to the electron spin or momentum. For the other terms one does need information about the electron spin or momentum. Apart from this division, we have put conventional terms that do not violate Lorentz symmetry in the front. Finally, we ordered the terms according to their need for the knowledge of the neutrino momentum (or recoil of the daughter nucleus), nuclear spin, or both, to be experimentally accessible.

\section{Contributions to $\chi^{\mu\nu}$} \label{sec:chicontrib}
In this section we discuss what kind of Lorentz-violating parameters in an effective field theory are accessible through a general Lorentz-violating tensor $\chi^{\mu\nu}$ that corrects the effective interaction Hamiltonian for $\beta$ decay as in Eq.~\eqref{betahamiltonian}. We consider two kinds of contributions to this tensor. Firstly, we discuss direct contributions to the propagator of the $W$-boson coming from the kinetic gauge sector and the Higgs-gauge sector of an effective field theory. Secondly, we discuss contributions coming from vertex corrections.

To discuss contributions to the $W$-boson propagator we first consider the Standard Model Extension (SME) due to  Colladay, Kosteleck\'y, and collaborators \cite{Colladay1997,Colladay1998a,Colladay1998b}. The SME is a general framework that in principle contains all effects of spontaneous Lorentz violation that can be incorporated into an effective field theory, while preserving all the symmetries of the SM except Lorentz and CPT symmetry. All terms up to mass dimension four are given explicitly in Ref.~\cite{Colladay1998b}. This renormalizable part of the SME we call the minimal SME (mSME). Lorentz-violating contributions to the $W$-boson propagator at tree level come from the gauge sector and the Higgs sector of the mSME. Restricting ourselves to these sectors, we can obtain an expression for the $W$-boson propagator to first order in Lorentz violation and in unitarity gauge. It is given by
\begin{multline}
\left\langle W^{\mu+}(p)W^{\nu-}(-p)\right\rangle = \frac{-i}{p^2-M_W^2}\left\{g^{\mu\nu} - \frac{p_\mu p_\nu}{M_W^2} + \frac{M_W^2}{p^2-M_W^2}\left(k_{\phi\phi}^{\mu\nu}+\tfrac{i}{2g}k_{\phi W}^{\mu\nu}\right) \right. \\ 
\left. - \frac{1}{p^2-M_W^2}\left(2k_W^{\rho\mu\sigma\nu}p_\rho p_\sigma + p^\mu p_\rho (k_{\phi\phi}^{\rho\nu}+\tfrac{i}{2g}k_{\phi W}^{\rho\nu})+p^\nu p_\rho(k_{\phi\phi}^{\rho\mu}+\tfrac{i}{2g}k_{\phi W}^{\rho\mu})\right) + \frac{k_{\phi\phi}^{\rho\sigma}p_\rho p_\sigma p_\mu p_\nu}{M_W^2(p^2-M_W^2)} \right\} \ .
\label{smewbosonpropagator}
\end{multline}
All coefficients that break Lorentz invariance are defined as in Ref. \cite{Colladay1998b}. The coupling constant $g$ that appears in the denominator of the $k_{\phi W}$ terms is the $SU(2)$ coupling contant. From this the relevant parameters for $\beta$ decay can be identified. Comparing the low-energy approximation of Eq.~\eqref{smewbosonpropagator} to Eq.~\eqref{wpropagator} we see that
\begin{equation}
\chi^{\mu\nu} = - k_{\phi\phi}^{\mu\nu} - \frac{i}{2g}k_{\phi W}^{\mu\nu} + \frac{2p_\rho p_\sigma}{M_W^2} k_W^{\rho\mu\sigma\nu} \ .
\label{lowenergysmeprop}
\end{equation}
Other terms are suppressed by powers of the four-momentum over $M_W$ relative to these.

Lorentz-violating terms of mass dimension higher than four will also give contributions to the propagator. In Ref.~\cite{Bolokhov2008} a classification is given of dimension-five Lorentz-violating operators that are irreducible to lower-dimensional operators by the equations of motion. As an example of a dimension-five contribution to $\chi^{\mu\nu}$, we consider the Lorentz-violating operator coming from the pure gauge sector 
\begin{equation}
\mathcal{L}_5 \supset C^{\mu\nu\rho}\; \mathrm{tr}W_{\mu\lambda}D_\nu \widetilde{W}_\rho^{\;\;\lambda} \ ,
\label{dim5bolokhov}
\end{equation}
where $W_{\mu\nu}$ is the $SU(2)$ gauge field strength and $\widetilde{W}_{\mu\nu}$ is its dual. The tensor $C^{\mu\nu\rho}$ is symmetric in all its indices and has mass dimension $-1$. (As discussed in Ref.~\cite{Kostel2009} for the photon, there are additional operators that contribute when the gauge boson is off-shell, as is the case for the $W$-boson in $\beta$ decay.) When including only the term in Eq.~\eqref{dim5bolokhov}, the $W$-boson propagator to first order and in unitarity gauge reads
\begin{equation}
\left\langle W^{\mu+}(p)W^{\nu-}(-p)\right\rangle = \frac{-i}{p^2-M_W^2}\left(g^{\mu\nu}-\frac{p^\mu p^\nu}{M_W^2}-\frac{4\epsilon^{\rho\mu\alpha\nu}C^{\lambda\beta}_{\;\;\;\;\rho}p_\lambda p_\beta p_\alpha}{(p^2-M_W^2)} \right) \ .
\label{LVdim5prop}
\end{equation}

Apart from the corrections in Eq.~\eqref{LVdim5prop} and the preceding discussion, there are also corrections from the dimension-five Higgs-gauge sector and higher-dimensional operators. Lorentz-violating operators of dimension six and higher have not been fully classified \cite{Mattingly2008,Liberati:2009pf}. However, the general tensor $\chi^{\mu\nu}$ in principle includes the effects of all Lorentz-violating contributions to the propagator. If all contributions to $\chi^{\mu\nu}$ come from the $W$-boson propagator directly ({\it i.e.} not from the vertex), we can write
\begin{equation}
\chi_p^{\mu\nu} = \sum_{n}^{\infty}Y_n^{\mu\nu\alpha_1\cdots\alpha_n}p_{\alpha_1}\cdots p_{\alpha_n} \ ,
\label{chimunup}
\end{equation}
where the subscript $p$ on $\chi_p^{\mu\nu}$ signals that we only consider contributions coming directly from the tree-level propagator. The mass dimension of $Y_n$ is $-n$, which implies that $\chi_p^{\mu\nu}$ is dimensionless. This does not necessarily mean that $Y_n$ is suppressed by $n$ powers of the mass scale at which Lorentz symmetry is broken ({\it e.g.} the Planck mass), since it can contain powers of the $W$-boson mass. From Eq.~\eqref{LVdim5prop}, for example, we obtain $Y_3^{\mu\nu\lambda\beta\alpha} = 4\epsilon^{\rho\mu\alpha\nu} C^{\lambda\beta}_{\;\;\;\rho}/M_W^2$, which is suppressed by one power of the Lorentz-breaking scale, through the $C$ coefficient of mass dimension $-1$. 

If we neglect terms that are suppressed by powers of the four momentum over the $W$-boson mass with respect to other terms containing the same Lorentz-violating parameter, hermiticity of the Lagrangian implies that
\begin{equation}
\chi_p^{\mu\nu*}(p) = \chi_p^{\nu\mu}(-p) \ .
\label{hermchi}
\end{equation}
This relation can be useful to limit the number of terms in Eq.~\eqref{finalbetaresult} for a particular Lorentz-violating operator. The dimension-five operator in Eq.~\eqref{dim5bolokhov} for example has $\chi^{\mu\nu*}=-\chi^{\nu\mu}$. Since the $\mu$ and $\nu$ indices are on the epsilon tensor, $\chi^{\mu\nu}$ is antisymmetric and real. This severely reduces the number of terms in Eq.~\eqref{LVquantities}. When the number of occurrences of $p$ in $\chi^{\mu\nu}_p$ is odd, CPT is violated. 

We now look at contributions to $\chi^{\mu\nu}$ coming from Lorentz-violating corrections to a vertex connecting a left-handed fermion current and a $W$-boson. In general, the vertex will have the form 
\begin{equation}
   -i\gamma_\nu(g^{\mu\nu} + \chi_v^{\mu\nu}) \ ,
\label{vertexcorr}
\end{equation}
where the subscript $v$ means ``vertex.'' When contracting this corrected vertex with a Lorentz-symmetric propagator at low energy, this will give the same correction to $\beta$ decay as Eq.~\eqref{wpropagator}. In the following, we will consider only the quark vertex, since parameters from the quark sector of the mSME have been much less constrained than those that contribute to the lepton vertex. The analysis is completely analogous, however.

We consider gauge-invariant terms that can contribute to the vertex containing a few ingredients. First of all, the terms contain the left-handed quark doublet $Q^i = \left(\begin{array}{c}u^i_L \\ d^i_L \end{array}\right)$, where the index $i$ runs over the three quark generations (although for $\beta$ decay we only need the first generation). Secondly, we consider covariant derivatives of the quark doublets $D^\mu Q$. And finally, the terms under consideration may contain the $SU(2)$ gauge field strength $W^{\mu\nu}$ and its covariant derivatives. However, the gauge field strength can only be present once, since otherwise the term no longer describes a three-point interaction. One could also use the Higgs doublet to build terms contributing to the vertex (see Ref. \cite{Bolokhov2008} for examples), but since our goal is to illustrate the generality of the use of $\chi^{\mu\nu}$, and not to give an exhaustive list of all terms contributing to it, we will settle for the previously mentioned ingredients. Using these ingredients, we can build two types of terms. The first type does not contain the gauge field strength and looks like
\begin{equation}
\mathcal{L}_{n+3} = T_{ij}^{\mu \alpha_1 \cdots \alpha_n}\bar{Q}_i \gamma_\mu (iD_{\alpha_1})\cdots (iD_{\alpha_n}) Q_j \ ,
\label{typeIvertex}
\end{equation}
where $n+3$ is the mass dimension of the operator and $T_{ij}$ is a hermitian matrix in generation space of mass dimension $1-n$. The gamma-matrix structure is limited to an odd number of gamma matrices, because the term is built out of a left-handed quark doublet and its conjugate. This was already mentioned in Ref.~\cite{Bolokhov2008} and it means that the gamma-matrix structure in Eq.~\eqref{typeIvertex} is exhaustive. As an example we mention that at mass dimension four the type of term as given in Eq.~\eqref{typeIvertex} is the only gauge-invariant term that gives a contribution to the vertex correction Eq.~\eqref{vertexcorr} for quarks. It is part of the mSME and was given in Ref.~\cite{Colladay1998b}. Adapted to our notation, this dimension-four term looks like
\begin{equation}
\mathcal{L}_4 \supset i(c_Q)_{ij}^{\mu\alpha}\bar{Q}_i\gamma_\mu D_\alpha Q_j \ .
\label{msmevertex}
\end{equation}

The second type of terms that contribute to Lorentz violation in the vertex contains one instance of the $SU(2)$ field strength or its covariant derivatives and looks like
\begin{equation}
\mathcal{L}_{m+n+5} = F_{ij}^{\mu\nu\rho\alpha_1\cdots\alpha_n\beta_1\cdots\beta_m}\bar{Q}_i \gamma_\mu \left[D_{\alpha_1}\cdots D_{\alpha_n} W_{\nu\rho}\right] (iD_{\beta_1}) \cdots (iD_{\beta_m})Q_j \ .
\label{typeIIvertex}
\end{equation}
Here $m+n+5$ is the mass dimension of the operator and $F_{ij}$ is again a hermitian matrix in generation space. An example of a term like this can be found in Ref. \cite{Bolokhov2008} and it looks like
\begin{equation}
\mathcal{L}_5 \supset (c_{Q,3})^{\mu}_{ij}\bar{Q}_i\gamma^\lambda W_{\mu\lambda}Q_j \ .
\end{equation}
It is clear from Eqs. \eqref{typeIvertex} and \eqref{typeIIvertex} that a general contribution to the parameter $\chi^{\mu\nu}$ coming from the vertex will have the form
\begin{equation}
\chi_v^{\mu\nu}  = \sum_{n=0}^\infty \sum_{m=0}^\infty V^{\mu\nu\alpha_1\cdots\alpha_n\beta_1\cdots\beta_m}p_{\alpha_1}\cdots p_{\alpha_n}q_{\beta_1}\cdots q_{\beta_m} \ ,
\end{equation}
where $p$ is the $W$-boson momentum and $q$ is the momentum of one of the quarks (the momentum of the other quark can be eliminated by using momentum conservation). Including terms containing the Higgs doublet will not change this form of the vertex contribution.

There is one problem with the first type of terms, given in Eq.~\eqref{typeIvertex}. These terms always contain a part that is a kinetic quark term. These kinetic terms should be taken into account in a calculation of $\beta$ decay. It is not clear, however, how such terms manifest themselves in effective parameters for the nuclei. Therefore, in principle we cannot fully treat the kind of terms given in Eq.~\eqref{typeIvertex} in $\beta$ decay, at least not in the way we discussed in the previous sections.

A way around this problem in some cases may be found in field redefinitions. As was shown in Ref.~\cite{Colladay:2002eh}, some Lorentz-violating parameters have no observable consequences and can be removed from the Lagrangian by suitable field redefinitions. Some parameters can be removed from a free-field theory, but as soon as interactions are included, the interaction terms prevent their removal. This means that they can be removed from the non-interacting part of the Lagrangian, but the field redefinition used to do this will make the interaction terms Lorentz non-invariant. What can be accomplished, therefore, is that these parameters are moved from the free-field equations to the interactions. However, this is not possible for all parameters and a full analysis of this issue lies outside the scope of this paper.

From the preceding considerations, it is clear that using $\chi^{\mu\nu}$ is a general approach to Lorentz violation in $\beta$ decay. Moreover, the propagator in Eq.~\eqref{wpropagator} is, to first order in Lorentz violation, compatible with the low-energy limit of the propagator of massive photons given in Ref.~\cite{cambiaso2012}; {\it cf.} also Ref.~\cite{Alt12}. It has become clear that $\chi^{\mu\nu}$ can depend on the $W$-boson or quark momenta. Although this does not matter for Eq.~\eqref{finalbetaresult}, the results after integrating over momenta (of the neutrino or the $\beta$ particle) will differ for different momentum dependences of $\chi^{\mu\nu}$. Since it is the dominant contribution, we will only consider a momentum-independent $\chi^{\mu\nu}$ when performing an integration in the following sections. In Appendix \ref{appendixchimomdep} we will work out an example where $\chi^{\mu\nu}$ depends on the momentum of the $W$-boson.

\section{Discussion} \label{sec:discussion}
In this section we explore the experimental signatures of Eq.~\eqref{finalbetaresult}.  We first discuss the difference between observer and particle Lorentz transformations, then we give expressions for decay rates that might be of interest to experiments, and finally we explore some dedicated experiments relevant for testing Lorentz violation in $\beta$ decay. A complete overview of the observable Lorentz-violating effects in different $\beta$-decay transitions (Fermi, Gamow-Teller, or mixed) is delegated to Appendix \ref{appendixobservables}, where a Table is given that contains the full set of Lorentz-violating vectors and tensors of Eq. \eqref{LVquantities} and their observability in different transitions.

\subsection{Particle and observer Lorentz transformations}
To understand what the expressions for the Lorentz-violating decay rates imply for experiments, it is important to differentiate between observer and particle Lorentz transformations \cite{Colladay1997}. All Lorentz-violating (tensor) coefficients in the SME can be viewed as constant background fields. In an ``observer Lorentz transformation'' one transforms all quantities, including the background fields. This is merely a coordinate transformation and the physics should be invariant under it. Indeed, observer Lorentz invariance is built into the SME. In practice, it means that all Lorentz indices are contracted. Doing a ``particle Lorentz transformation,'' however, transforms all quantities, except the background fields. This corresponds to physically boosting or rotating the experiment with respect to the background. This happens, for example, when the Earth moves through space and the experiment moves with it. As seen from the laboratory frame, the values of $\chi^{\mu\nu}$ will change as the Earth changes its orientation. These values will thus oscillate with the rotation frequency of the Earth. Because {\it a priori} there is no preferred frame (except perhaps that defined by the cosmic microwave background), we have to choose a standard reference frame. In this frame we then compare bounds on Lorentz-violating parameters from different experiments. The customary choice is the Sun-centered inertial reference frame \cite{Kostelecky2011}. We have to express the Lorentz-violating quantities in Eq.~\eqref{LVquantities}, which are given in the laboratory frame (denoted by $\chi^{\mu\nu}$), in terms of Lorentz-violating coefficients in this Sun-centered inertial reference frame (denoted by $X^{\mu\nu}$). This is accomplished by using a rotation matrix $R$:
\begin{equation}
\chi^{\mu\nu} = R^{\mu}_{\;\;\rho}R^{\nu}_{\;\;\sigma}X^{\rho\sigma} \ ,
\label{frametrafo}
\end{equation}
with $R$ given by
\begin{equation}
R(\zeta,t) = \left(
\begin{array}{cccc}
1 & 0 & 0 & 0 \\
0 & \cos\zeta \cos\Omega t & \cos\zeta \sin\Omega t & - \sin\zeta \\
0 & -\sin\Omega t & \cos\Omega t & 0 \\
0 & \sin\zeta\cos\Omega t & \sin\zeta\sin\Omega t & \cos\zeta
\end{array}
\right) \ ,
\label{rotationmatrix}
\end{equation}
where $\zeta$ is the colatitude of the experiment and $\Omega\simeq 2\pi/(23\mathrm{h}\,56\mathrm{m})$ is the Earth's sidereal rotation frequency. This form of $R$ is correct if the laboratory coordinate axes $(\hat{x},\hat{y},\hat{z})$ are defined such that $\hat{z}$ is perpendicular to the Earth's surface, $\hat{x}$ points in the north-south direction, and the laboratory $\hat{y}$-direction completes the right-handed coordinate system by pointing from west to east. Further details on the Sun-centered inertial reference frame are given in Ref.~\cite{Kostelecky2011} and references therein. 

It is now easy to see that the parts of Eq.~\eqref{finalbetaresult} that depend on the $\chi^{0j}$ or $\chi^{j0}$, with $j$ a space-like index, will oscillate with a frequency equal to the rotation frequency of the Earth, while quantities depending on $\chi^{jk}$ in addition have parts that oscillate with twice this frequency. Because Eq.~\eqref{rotationmatrix} contains no boost, the isotropic coefficient $\chi^{00}$ is the same in frames related by a rotation. This parameter is thus only accessible if one measures the decay rate at velocities high enough for the relativistic gamma factor to become significant. For Earth-bound neutron and nuclear $\beta$-decay experiments $\chi^{00}$ is therefore irrelevant, but it could affect accelerator-based experiments and cosmic rays.

\subsection{Sample decay rates}
In this section we work out some key examples of decay rates, a pure Fermi transition, a pure Gamow-Teller transition, and neutron decay (an example of a mixed transition). For the neutron we will work out the decay rate in terms of Lorentz-violating coefficients, given in the Sun-centered intertial reference frame, to explicitly show the oscillations with time due to the rotation of the Earth.

We start with a pure Fermi transition, which has $\left\langle \sigma \right\rangle=0$. Integrating over neutrino energy and direction and summing over electron polarization gives
\begin{equation}
dW_F = dW^0\left[1+2\chi_r^{00}+\frac{2\chi_r^{0l} p^l}{E_e}\right] \ ,
\label{fermitransition}
\end{equation}
with
\begin{equation}
dW^0 = \frac{1}{8\pi^4}|{\bf p}|E_e(E_e-E_0)^2dE_e d\Omega_e F(E_e,\pm Z)\xi \ .
\end{equation}
Therefore, measuring the $\beta$-decay rate in different directions gives access to $\chi_r^{0l}$.

For a Gamow-Teller transition in randomly oriented nuclei, after summing over electron polarization and integrating over neutrino energy and momentum, we get
\begin{equation}
dW_{GT} = dW^0\left[1-\tfrac{2}{3}\chi_r^{00} + \tfrac{2}{3}(\chi_r^{l0} + \tilde{\chi}_i^l)\frac{p^l}{E_e}\right] \ .
\label{gttransitionrandom}
\end{equation}
Measuring Fermi transitions and Gamow-Teller transitions gives access to different parameters. In the former one measures $\chi_r^{0l}$, while in the latter bounds on a combination of $\chi_r^{l0}$ and $\tilde{\chi}_i^{l}$ are possible. In the case that $(\chi^{\mu\nu})^* = \chi^{\nu\mu}$, it follows that $\chi_r^{0l}=\chi_r^{l0}$, so one can disentangle the bounds on $\chi_r^{l0}$ and $\tilde{\chi}_i^{l}$.

For other parts of $\chi^{\mu\nu}$ it is necessary to have more directional information from the experiment. One way to accomplish this is by polarizing the parent nuclei. The relevant expression for the Gamow-Teller transition rate of polarized nuclei is
\begin{eqnarray}
dW_{GT} &=& dW^0\left[1-\tfrac{2}{3}\chi_r^{00} + \tfrac{2}{3}(\chi_r^{l0} + \tilde{\chi}_i^l)\frac{p^l}{E_e}\right] \notag \\
&& \mp \Lambda^{(1)}\left[(1-\chi_r^{00})\frac{{\bf p \cdot \hat{I}}}{E_e} + \tilde{\chi}_i^l \Ip^l + \frac{\chi_r^{lk}p^l \Ip^k}{E_e}-\frac{\chi_r^{l0}({\bf p}\times {\bf \hat{I}})^l}{E_e}\right] \notag \\
&& + \Lambda^{(2)}\left[-\chi_r^{00}+(\chi_r^{l0}+\tilde{\chi}_i^l)\frac{p^l}{E_e}+3\chi_r^{kl}\Ip^k \Ip^l - 3\chi_r^{l0}\Ip^l \frac{{\bf p \cdot \hat{I}}}{E_e} - 3\chi_i^{ml}\Ip^m\frac{({\bf p}\times {\bf \hat{I}})^l}{E_e}\right] \ .
\end{eqnarray}
By combining, for example, an asymmetry measurement for spin up and spin down, with measurements of Eqs.~\eqref{fermitransition} and \eqref{gttransitionrandom}, one can extract bounds on $\chi_r^{lk}$, $\chi_r^{0l}$, and $\tilde{\chi}_i^{l}$ (again assuming $(\chi^{\mu\nu})^* = \chi^{\nu\mu}$).
In Appendix \ref{appendixobservables} we relate $\chi^{\mu\nu}$ to observables in Fermi, Gamow-Teller, and mixed transitions with respect to the decay parameters in Eqs.~(\ref{finalbetaresult}) and (\ref{LVquantities}).

As a final example we look at polarized neutron $\beta$ decay. We assume that one measures the direction and energy of the outgoing electron. An experiment relevant for this example is described in Ref.~\cite{KozelaCPT}. For the neutron, the ratio of $C_V$ and $C_A$ has been experimentally determined to be $C_A/C_V = -1.27$. Since $\left\langle 1\right\rangle =1$ and $\left\langle \sigma \right\rangle = \sqrt{3}$ for neutron decay, it follows that $x=0.17$ and $y=-0.37$. By using these values, the differential decay rate of polarized neutrons becomes
\begin{eqnarray}
dW & = & dW^0 \left\{1 - 0.21\chi_r^{00} + (0.34 \chi_r^{0l} + 0.55(\chi_r^{l0} + \tilde{\chi}_i^{l}))\frac{p^l}{E_e}\right. \notag \\
&& +\left.\frac{\left\langle m \right\rangle}{j}\Ip^k\left[0.43(\chi_r^{k0}-\chi_r^{0k})-0.55\tilde{\chi}_i^k - (0.12-0.99\chi_r^{00})\frac{p^k}{E_e}-0.99(\chi_r^{lk}-\chi_i^{s0}\epsilon^{ksl})\frac{p^l}{E_e}\right]\right\} \ . \notag \\
\label{neutrondecayrate}
\end{eqnarray}
There are no second-order anisotropic contributions for the neutron, because $\Lambda^{(2)}=0$ for particles with spin-$\tfrac{1}{2}$.

To determine the Lorentz-violating decay rate of the neutron in terms of coefficients defined in the Sun-centered inertial reference frame we use Eq.~\eqref{frametrafo}. We take the quantization axis of the nuclear spin to be in the $+\hat{z}$ direction and we assume that the electrons are measured in the $\hat{x}$-$\hat{z}$ plane and that ${\bf p \cdot \hat{I}}= v E_e \cos\theta$, with $v=|{\bf p}|/E_e$ the velocity of the electron, while $\mathcal{P}$ is the polarization of the parent nuclei. The resulting expression reads
\begin{eqnarray}
dW  &=& dW^0\left[1 - 0.12 \mathcal{P}v \cos\theta - X_r^{TT}(0.21-0.99 \mathcal{P}v\cos\theta)\right. \notag \\ 
&&\left. + Z_1 + Z_2 \cos(\Omega t) + Z_3 \sin(\Omega t) + Z_4 \cos(2\Omega t) + Z_5 \sin(2\Omega t)\right] \ ,
\label{finalneutron}
\end{eqnarray}
where the quantities $Z_i$, depending on the colatitude of the experiment, are given by
\begin{subequations}
\begin{eqnarray}
Z_1 & = & v(0.55(X_i^{XY}-X_i^{YX}+X_r^{ZT})+0.34X_r^{TZ})\cos(\theta+\zeta)+\notag \\
&& + \mathcal{P}\left[(0.55(X_i^{YX}-X_i^{XY})+0.43(X_r^{ZT}-X_r^{TZ}))\cos\zeta \right.\notag \\\
&& \left. - 0.99 v X_r^{ZZ}\cos(\theta+\zeta)\cos\zeta - 0.50 v(X_r^{XX}+X_r^{YY})\sin(\theta+\zeta)\sin\zeta \right] \ , \\
Z_2 & = & 0.55 v ((X_i^{YZ}-X_i^{ZY}+X_r^{XT})+0.34X_r^{TX})\sin(\theta+\zeta) \notag \\
&& +\mathcal{P}\left[(0.55(X_i^{ZY}-X_i^{YZ}) + 0.43(X_r^{XT}-X_r^{TX}))\sin\zeta \right.  \notag \\
&& -0.99 v (X_r^{XZ}+X_r^{ZX})\cos\theta\sin\zeta\cos\zeta  \notag \\
&& \left. - 0.99 v (X_i^{YT}+X_r^{XZ}\cos^2\zeta - X_r^{ZX}\sin^2\zeta)\sin\theta\right] \ , \\
Z_3 & = & 0.55 v ((X_i^{ZX}-X_i^{XZ}+X_r^{YT})+0.34X_r^{TY})\sin(\theta+\zeta) \notag \\
&& +\mathcal{P}\left[(0.55(X_i^{XZ}-X_i^{ZX}) + 0.43(X_r^{YT}-X_r^{TY}))\sin\zeta \right. \notag \\
&& -0.99 v (X_r^{YZ}+X_r^{ZY})\cos\theta\sin\zeta\cos\zeta \notag \\
&& \left. + 0.99 v (X_i^{XT}-X_r^{YZ}\cos^2\zeta + X_r^{ZY}\sin^2\zeta)\sin\theta\right] \ , \\
Z_4 & = & 0.50 v \mathcal{P} (X_r^{YY}-X_r^{XX})\sin(\theta+\zeta)\sin\zeta \ , \\
Z_5 & = & -0.50 v \mathcal{P} (X_r^{XY}+X_r^{YX})\sin(\theta+\zeta)\sin\zeta \ .
\end{eqnarray}
\end{subequations}
We see indeed that Eq.~\eqref{finalneutron} has parts ($Z_2$ and $Z_3$) that oscillate with a period of one sidereal day  and it has parts ($Z_4$ and $Z_5$) that have half this oscillation period. 

\subsection{Relevant experiments}
To put bounds on $\chi^{\mu\nu}$ it is necessary to measure a direction, {\it i.e.} ${\bf \hat{p}}$, $\hat{I}$, $\hat{k}$, or ${\bf \hat{s}_e}$. The only exception is the purely timelike part of $\chi^{\mu\nu}$. To get a limit on $\chi^{00}$ one would have to compare experiments that have relative velocities large enough to get sizable relativistic gamma factors. It is, in general, necessary to measure over longer periods of time and record data with ``time-stamps.'' In this way, one can search for oscillatory effects with periods of one or one-half sidereal day, which arise in the case of Lorentz violation. 

These considerations imply that almost all $\beta$-decay experiments that have been performed in the past are irrelevant to the question of Lorentz violation. In the 1970s, however, two experiments were performed to test the rotational invariance of $\beta$ decay \cite{Newman,Ullman}. These experiments were done by using nuclei that have ``forbidden'' $\beta$-decay modes \cite{noordmansinprep}, assuming Lorentz violation would manifest itself in angular-momentum violation. As mentioned earlier, at PSI in Switzerland, experiments on the $\beta$ decay of polarized neutrons were performed. The data was time-stamped and an analysis looking for Lorentz invariance violation was performed \cite{KozelaCPT}. In the light of our result for neutron decay in Eq. \eqref{neutrondecayrate} we believe that a more complete analysis could be performed, from which bounds on Lorentz-violating parameters in the SME could be obtained. Finally, we mention the ongoing experiment performed at our accelerator facility KVI in Groningen, which will search for Lorentz-violating effects in the Gamow-Teller $\beta$ decay of $^{20}$Na \cite{kviexperiment}. This experiment is designed to be sensitive to Lorentz-violating effects arising from first-order anisotropic terms in Eq. \eqref{finalbetaresult} by controlled flipping of the spin of the parent nucleus.  The theoretical framework developed in this paper will be used in the analysis and interpretation of this experiment.

\section*{Acknowledgments}
We thank our colleagues at KVI, in particular Stefan M\"uller and Gerco Onderwater, for discussions,
and Alan Kosteleck\'y for a useful communication.
JN thanks Alan Kosteleck\'y and Ralf Lehnert for helpful discussions, and the organizers of the IUCSS
Summer School on the SME in 2012 for their hospitality. This research was supported by the Dutch
Stichting voor Fundamenteel Onderzoek der Materie (FOM) under Programmes 104 and 114 and
project 08PR2636.

\appendix

\section{Lepton currents} \label{lepcurrentsapp}
Omitting the factor $F(E_e,\pm Z)$ and the subscript $\mp$ on $P_{\mp}$, the different combinations of lepton currents are given by
\begin{eqnarray}
\tfrac{1}{8}|J^0|^2 & = & P^0\left(\tfrac{1}{2}(1+2\chi_r^{00})k^0-\chi_r^{0l}k^l\right) - P^l\left(-\tfrac{1}{2}(1+2\chi_r^{00})k^l-\chi_r^{0l}k^0 + \chi_i^{0j}\epsilon^{jml}k^m\right) \ ,
\end{eqnarray}
\begin{eqnarray}
\tfrac{1}{8}(|J_{+1}|^2+|J_{-1}|^2+|J_z|^2) & = & P^0\left((\tfrac{3}{2}-\chi_r^{00})k^0 + (\chi_r^{l0}-\tilde{\chi}_i^l)k^l\right) \notag \\
&& - P^l\left((\tfrac{1}{2}-\chi_r^{00})k^l-(\chi_r^{l0}+\tilde{\chi}_i^l)k^0 + (\chi_r^{ml}+\chi_r^{lm}+\chi_i^{j0}\epsilon^{ljm})k^m\right) \ , \notag \\
\end{eqnarray}
\begin{eqnarray}
\tfrac{1}{8}(|J_{+1}|^2-|J_{-1}|^2) & = & \pm P^0\left((1-\chi_r^{00})({\bf k\cdot \hat{I}}) + \chi_i^{l0}({\bf k\times\hat{I}})^l - k^0( \boldsymbol{\tilde{\chi}_i} \cdot { \bf \hat{I} }) + \chi_r^{lm}k^l \Ip^m\right) \notag \\
&& \mp P^l \left((1-\chi_r^{00})k^0\Ip^l - k^0\epsilon^{mkl}\chi_i^{k0}\Ip^m + \chi_r^{m0}k^m \Ip^l - k^l (\boldsymbol{\tilde{\chi}_i}\cdot {\bf \hat{I} }) - \chi_r^{l0}({\bf k\cdot \hat{I}}) + k^0 \chi_r^{lm}\Ip^m  \right. \notag \\
&& \left.+ \chi_i^{ml}({\bf k\times \hat{I}})^m + \epsilon^{ljm}\chi_i^{mk}k^k \Ip^j \right) \ ,
\label{JplusminJmin}
\end{eqnarray}
\begin{eqnarray}
\tfrac{1}{8}(|J_{+1}|^2+|J_{-1}|^2-2|J_z|^2) & = & P^0\left((3\chi_r^{lm}\Ip^l\Ip^m - \chi_r^{00})k^0 + (\chi_r^{l0}-3\chi_r^{m0}\Ip^m \Ip^l - \tilde{\chi}_i^l)k^l + 3\chi_i^{ml}\Ip^m({\bf k\times \hat{I}} )^l\right) \notag \\
&& -P^l\left(3({\bf k\cdot\hat{I}})\Ip^l -(\chi_r^{l0}+\tilde{\chi}_i^l-3\chi_r^{m0}\Ip^m\Ip^l-3\chi_i^{mj}\epsilon^{jlk}\Ip^m\Ip^k )k^0  \right. \notag \\
&& - (1+\chi_r^{00}-3\chi_r^{mj}\Ip^m\Ip^j)k^l  + 3\chi_i^{m0}\Ip^m ({\bf k\times\hat{I}})^l \notag \\
&& \left.+ (\chi_r^{ml}+\chi_r^{lm}+\chi_i^{j0}\epsilon^{ljm}-3\chi_r^{jl}\Ip^j\Ip^m-3\chi_r^{jm}\Ip^j\Ip^l)k^m \right) \ ,
\end{eqnarray}
\begin{eqnarray}
\tfrac{1}{8}(J^0 J^*_z+J_z J^{0*}) & = & P^0\left((1+\chi_r^{00})({\bf k\cdot\hat{I}})+(\chi_r^{m0}-\chi_r^{0m})\Ip^m k^0-\chi_r^{ml}\Ip^m k^l -\chi_i^{0l}({\bf k\times \hat{I}})^l\right) \notag \\
&& -P^l\left(-(1+\chi_r^{00})\Ip^l k^0 +\chi_r^{0l}({\bf k\cdot\hat{I}})- (\chi_r^{0m}+\chi_r^{m0})\Ip^m k^l + \chi_r^{0m}k^m \Ip^l  \right. \notag \\
&&\left. +\chi_r^{ml}\Ip^m k^0 - \chi_i^{0k}\epsilon^{mkl}\Ip^m k^0 + \chi_i^{mj}\epsilon^{jkl}k^k\Ip^m -\chi_i^{00}({\bf k\times\hat{I}})^l\right) \ ,
\end{eqnarray}
where the subscripts $r$ and $i$ label the real and imaginary parts of $\chi^{\mu\nu}$, while $\tilde{\chi}^j = \epsilon^{jmk}\chi^{mk}$, the upper (lower) signs in Eq.~\eqref{JplusminJmin} refer to the case of the electron (positron), the Latin indices run over the spatial components $1,2,3$, and summation over repeated indices is understood.

\section{Momentum-dependent $\chi^{\mu\nu}$} \label{appendixchimomdep}
In section \ref{sec:chicontrib} we noted that it is possible for $\chi^{\mu\nu}$ to depend on the $W$-boson momentum or on the momentum of one of the quarks. We can just substitute any form of $\chi^{\mu\nu}$ in Eq.~\eqref{finalbetaresult}, but the integrals over momentum will become more complicated when $\chi^{\mu\nu}$ depends on these momenta. In particular, we will get integrals over the angular coordinates of the momentum. A useful relation in doing such integrals is
\begin{equation}
\int d\Omega\ T^{i_1\cdots i_n}\hat{k}_{i_1}\cdots \hat{k}_{i_n} = \left\{
\begin{array}{cr} 
\frac{4\pi}{(n+1)!!}T^{i_1\cdots i_n}(\delta_{i_1 i_2}\cdots\delta_{i_{n-1}i_n}+ \mathrm{all\ possible\ contractions}) & (n\ \mathrm{even}) \\
0 & (n\ \mathrm{odd}) 
\end{array} \ ,
\right.
\label{unitvintegral}
\end{equation}
where $T$ is an arbitrary tensor with $n$ indices, $\hat{k} = (\sin\theta\cos\phi,\sin\theta\sin\phi,\cos\theta)$, and all the $i$ indices are spatial indices running from $1$ to $3$.

As an example we work out the result for Fermi decay $(x=1,\;y=0)$ with $\chi^{\mu\nu}$ coming from the CPT even gauge sector of the mSME. In that case $\chi^{\mu\nu}$ can be read of from Eq.~\eqref{lowenergysmeprop} and is given by
\begin{equation}
\chi^{\mu\nu} = \frac{2q_\rho q_\sigma}{M_W^2}k_W^{\rho\mu\sigma\nu} \ ,
\end{equation}
where $q$ is now the $W$-boson momentum, which due to momentum conservation is equal to the sum of the momenta of the $\beta$ particle and the neutrino. Summing over electron polarization and noting that in this case $\chi_i^{\mu\nu}=0$, the Fermi decay rate can be written as
\begin{equation}
dW_F = \frac{dW^0 d\Omega_{\nu}}{4\pi}\left[\left(1+\frac{\bf p\cdot \hat{k}}{E_e}\right)\left(1+2\chi_r^{00}\right) + 2\chi_r^{0l}\left(\frac{p^l}{E_e}-\hat{k}^l\right)\right] \ .
\end{equation}
Performing the integral over the neutrino angles will only give a nonzero result if there is an even number of unit vectors in the direction of the neutrino momentum. By using Eq.~\eqref{unitvintegral} we find that the decay rate integrated over the neutrino momentum is given by
\begin{equation}
dW_F = dW^0\left[1+\frac{4}{3M_W^2}\left(k_W^{i0i0}E_e(E_e-E_0)+k_W^{i0ij}\frac{E_0(E_0-E_e)p^j}{E_e}+k_W^{i0j0}\frac{(E_e+5E_0)p^ip^j}{E_e}\right)\right] \ ,
\end{equation}
with summation over repeated indices implied, as usual.

\section{Experimental sensitivities} \label{appendixobservables}
In Table \ref{tab:obstable} we give an overview of the observability of the Lorentz-violating parameter
$\chi^{\mu\nu}$ in the different types of $\beta$-decay transitions, Fermi, Gamow-Teller, or mixed.
Although Table \ref{tab:obstable} contains the same information as Eqs.~\eqref{finalbetaresult} and
\eqref{LVquantities}, it provides an clearer overview of which Lorentz-violating parameters are
accessible in which kind of experiments.

\begin{table}[h]
\centering
\begin{tabular}{||c|c||ccccc|ccc|p{12em}||}
\hline
& & $\ccr^{00}$ & $\ccr^{0l}$ & $\ccr^{ml}$ & $\cci^{0l}$ & $\cci^{ml}$ & $\cci^{00}$ & $\ccr^{l0}$ & $\cci^{l0}$ & Comments  \\ \hline
t & & X & & & & & & & & \\ 
$w_1$ & $\hat{k}$ & & F & & & GT & & GT & & {\footnotesize $\cci^{(ml)}$ not accessible. } \\
$w_2$ & $\hat{I}$ & & M & & & GT & & & M & {\footnotesize If $\ccr^{[\mu\nu]}=0$, $\ccr^{0k}$ cancels $\ccr^{k0}$. $\cci^{(ml)}$ not accessible.}\\
$w_3$ & $\hat{p}$ & & F & & & GT & & GT & & {\footnotesize $\cci^{(ml)}$ not accessible. } \\
$T_1$ & $\hat{I}^*$ & & & GT & & & & & & {\footnotesize Vanishes for $j=\tfrac{1}{2}$} \\
$T_2$ & $\hat{I},\hat{k}$ & GT & & GT & M & & & & GT & \\
$T_3$ & $\hat{p},\hat{k}$ & X & & GT & F & & & & GT & {\footnotesize $\ccr^{[ml]}$ not accessible. } \\
$T_4$ & $\hat{p},\hat{I}$ & GT & & GT & M & & & & GT & \\
$S_1$ & $\hat{I}^*,\hat{k}$ & & & & & GT & & GT & & {\footnotesize Vanishes for $j=\tfrac{1}{2}$} \\
$S_2$ & $\hat{p},\hat{k}$ & & & & & GT & & GT & & {\footnotesize Vanishes for $j=\tfrac{1}{2}$} \\
$S_3$ & $\hat{p},\hat{I},\hat{k}$ & & M & & & GT & M & GT & &  \\
$R$ & $\hat{p},\hat{I}^*,\hat{k}$  & & & GT & & & & & GT & {\footnotesize Vanishes for $j=\tfrac{1}{2}$} \\
\hline
\end{tabular}
\caption{Observability of the Lorentz-violating parameter $\chi^{\mu\nu}$ in different $\beta$-decay transitions.}
\label{tab:obstable}
\end{table}

On the top line of Table \ref{tab:obstable} are the different parts of $\chi^{\mu\nu}$ that are independent under the action of rotations, in particular Eq.~\eqref{frametrafo}. If we assume that $\chi^{\mu\nu*}=\chi^{\nu\mu}$, the last three columns, corresponding to $\chi_i^{00}$, $\chi_r^{l0}$, and $\chi_i^{l0}$, are redundant. The first of these three columns should then be removed, since $\chi_i^{00}=0$ if we assume $\chi^{\mu\nu*}=\chi^{\nu\mu}$. In addition, the entries in the columns of $\chi_r^{l0}$ and $\chi_i^{l0}$ should be added to the columns of $\chi_r^{0l}$ and $\chi_i^{0l}$, respectively.

In the left-most column the Lorentz-violating quantities which are defined in Eq.~\eqref{LVquantities} are listed. In the second column we give the minimal directional information one needs to be able to see effects of the corresponding Lorentz-violating vector or tensor. Notice that the polarization of the $\beta$ particle does not appear. This is because it can be seen from Eq.~\eqref{finalbetaresult} that by measuring the electron direction one has access to the same Lorentz-violating observables as by measuring the polarization of the $\beta$ particle. An asterisk on $\Ip$ means that it occurs twice in the corresponding term in Eq.~\eqref{finalbetaresult}, and therefore it does not change sign if one flips the nuclear spin.
The entries in the Table show what is the relevant $\beta$-decay transition. F means that the component of $\chi^{\mu\nu}$ occurs in the Lorentz-violating quantity on the left in a Fermi or a mixed transition. GT has the analogous meaning for a Gamow-Teller or mixed transition. X means that the part of $\chi^{\mu\nu}$ on the top of the column shows up in the quantity in the first column in every kind of transition, while an M signals that it only is visible in a mixed transition.
Finally, we used the notations $\chi^{(\mu\nu)} = \tfrac{1}{2}(\chi^{\mu\nu}+\chi^{\nu\mu})$ and $\chi^{[\mu\nu]} = \tfrac{1}{2}(\chi^{\mu\nu}-\chi^{\nu\mu})$ in the right-most Comments column.


\end{document}